# The Relationship of the Laplacian Gauge to the Landau Gauge

Jeffrey E. Mandula

Department of Energy, Division of High Energy Physics
Washington, DC 20585, United States

The Laplacian gauge for gauge group $SU(N)$ is discussed in perturbation theory. It is shown that to the lowest non-trivial order, $\mathcal{O}(g^1)$, configurations in the Laplacian gauge automatically satisfy the (finite difference) Landau gauge condition. Laplacian gauge fixed configurations are examined numerically and it is seen that to $\mathcal{O}(g^2)$ they do not remain in the Landau gauge.

## 1. INTRODUCTION

In the continuum, the Landau gauge is normally formulated as a differential condition. There are Gribov copies[1], of course, and these are seen on the lattice as well. On the lattice one normally formulates the Landau gauge condition as a maximization.

$$\underset{g}{Max} \sum_{sites\ x} Re\ Tr \sum_\mu [U_\mu^g(x) + U_\mu^{g\dagger}(x-\hat{\mu})] \quad (1)$$
$$U_\mu^g(x) = g(x) U_\mu(x) g^\dagger(x+\hat{\mu})$$

The reason for this is not so much to deal with Gribov copies as to eliminate a plethora of lattice artifact copies. These copies have the property that at each site the quantity to be maximized in Eq. (1) could be either a maximum, a minimum, or even a saddle. The global maximization condition defines a copy free gauge in principle, but there is no algorithm for finding such a global maximum. There are generically many local maxima, which are the lattice versions of the continuum Gribov copies.

In 1992, Vink and Wiese proposed a copy free gauge that could be implemented on the lattice, which they called the Laplacian gauge[2]. It shares with the Landau gauge the virtue of smoothness, so that it naturally relates gauge fields and other gauge variant quantities at a distance. Additional developments of this gauge have included a compact perturbative formulation of the Laplacian gauge for $SU(2)$[3], computations of the quark and gluon propagators[4], and studies of instantons[5].

In this talk, I will perturbatively apply Vink and Wiese's gauge fixing algorithm to generic configurations for an $SU(N)$ or $U(N)$ lattice gauge theory. Two principal results will be presented regarding the relationship between the Laplacian and the Landau gauges. One is that in the lowest non-trivial order, $g^1$, configurations in the Laplacian gauge automatically satisfy the (finite difference) Landau gauge condition. The other is that ensembles of configurations numerically fixed to Laplacian gauge show a violation of the Landau gauge condition proportional to $g^2$.

## 2. THE LAPLACIAN GAUGE

The Laplacian gauge is defined in terms of the lattice Laplacian (color indices suppressed)

$$\Delta(x,y) = \sum_\mu [2\,\delta_{x,y} I - \delta_{x,y-\hat{\mu}} U_\mu(x) - \delta_{x-\hat{\mu},y} U_\mu^\dagger(y)] \quad (2)$$

$\Delta$ is a positive Hermitean operator. For $SU(N)$, one forms an $N{\times}N$ matrix field from the eigenvectors with the $N$ lowest eigenvalues,

$$M_{ji}(x) = (f_i(x))_j \quad (3)$$

performs a polar decomposition at each site, and defines a gauge transformation $\Omega(x)$ as the adjoint of the unitary part of $M(x)$. Applying this gauge transformation to the lattice link variables transforms the lattice to the Laplacian gauge. Vink and Wiese showed that the Laplacian gauge was free of Gribov copies, that is, the transformation to Laplacian gauge gave the same result, wherever one started on the gauge orbit of a given configuration. This does not violate Singer's theorem[6] that no continuous (in the field variables) gauge could be unique. There are discontinuities in the Laplacian gauge; they occur at configurations where $\Delta(x,y)$ has degeneracies among its first $N$ eigenvalues, which signals level crossing.

## 3. LATTICE PERTURBATION THEORY

Let us expand the Laplacian gauge algorithm in a power series in the bare lattice coupling. Define a lattice gauge potential by

$$U_\mu(x) = e^{-igA_\mu(x)} \quad (4)$$

and use the power series for $U_\mu$ to expand the Laplacian

$$\Delta(x,y) = \Delta^{(0)}(x,y) + g\Delta^{(1)}(x,y) + g^2\Delta^{(2)}(x,y) + \ldots \quad (5)$$

In zeroth order, i.e., $U_\mu(x) = I_N$, the Laplacian has an $N$-fold degenerate zero eigenvalue, which, since $\Delta^{(0)}$ is also a positive operator, are the lowest eigenvalues. Their eigenfunctions are the $N$ linearly independent constant $N$-vectors, which we denote as

$$f_i^{(0)}(x) = \xi_i^{(0)} \qquad i = 1,2,\ldots,N \quad (6)$$

We call the matrix formed from them $\Xi^{(0)}$.

From the orthogonality of the $f_i^{(0)}(x)$ and their constancy in $x$ we can choose their normalizations so that $\Xi^{(0)}$ is unitary. The degeneracy makes it easy to develop the analysis, but it also means that the perturbative expansion of the Laplacian gauge is a case of degenerate perturbation theory, in which higher order effects determine which linear combinations of the $f_i^{(0)}$ functions are the proper zeroth order eigenfunctions, i.e. remain unmixed in higher order. This higher order information, which in fact appears in second order, is needed even to correctly construct the zeroth order Laplacian gauge transformation[1].

Through second order, the equations for the eigenfunctions are

$$\Delta^{(0)} f_i^{(1)} + \Delta^{(1)} f_i^{(0)} = \lambda_i^{(1)} f_i^{(0)}$$
$$\Delta^{(0)} f_i^{(2)} + \Delta^{(1)} f_i^{(1)} + \Delta^{(2)} f_i^{(0)} = \lambda_i^{(1)} f_i^{(1)} + \lambda_i^{(2)} f_i^{(0)} \quad (7)$$

From the first equation it is seen that the first order shift in the eigenvalues vanishes, and the first order wave functions are a simple linear transformation of the zeroth order ones. This gives for the first order matrix $M$

$$M^{(1)}(x) = -i\tilde{\Delta}^{(0)-1}(x,y)\Delta_\mu A_\mu(y)\Xi^{(0)}$$
$$\equiv l_1(x)\Xi^{(0)} \quad (8)$$

where

$$\Delta_\mu A_\mu(y) = \sum_\mu [A_\mu(y) - A_\mu(y-\hat{\mu})] \quad (9)$$

is the lattice divergence of the gauge potential based on Eq.(4), and $\tilde{\Delta}^{(0)-1}$ is the inverse of the free Laplacian restricted to the space orthogonal to the $f_i^{(0)}(x)$.

Projecting the second order equation on the space of constant functions gives an ordinary $N \times N$ matrix eigenvalue problem:

$$P^{(0)}\left(\Delta^{(2)} - \Delta^{(1)}\tilde{\Delta}^{(0)-1}\Delta^{(1)}\right)f_i^{(0)} = \lambda_i^{(2)} f_i^{(0)} \quad (10)$$

Its solutions, which break the degeneracy between the $N$ lowest eigenvalues, are the stable zeroth order solutions. The orthogonal projection of this equation gives the second order term in $M$:

$$M^{(2)}(x) = -\tilde{\Delta}^{(0)-1}\left(\Delta^{(2)} - \Delta^{(1)}\tilde{\Delta}^{(0)-1}\Delta^{(1)}\right)\Xi^{(0)}$$
$$\equiv l_2(x)\Xi^{(0)} \quad (11)$$

These operator exercises are easily carried out in momentum space. The further construction of the transformation to Laplacian gauge proceeds one site at a time in coordinate space.

It is straightforward to check that $l_1(x)$ is antihermitian, so that to first order $M(x)$ is a unitary matrix at each site. However, at second order its unitary part must be explicitly extracted, giving through second order

$$\Omega^\dagger(x) = M(x)[M^\dagger(x)M(x)]^{-1/2}$$
$$\equiv \left(I_N + gl_1 + \frac{g^2}{2}(l_2^\dagger - l_2 + l_1^2)\right)\Xi^{(0)} \quad (12)$$

---

[1]In the case of $SU(2)$, the degeneracy persists to all orders because of conjugation symmetry, i.e., because the fundamental representation of $SU(2)$ is unitarily equivalent to its complex conjugate. This allows one to choose the matrix $M(x)$ proportional to the unit matrix[3].

**3.1 Transformation of the gauge potential**

It is worthwhile to express the transformation to Laplacian gauge in terms of the gauge potential, using the expansion of

$$U_\mu^\Omega(x) = e^{-igA_\mu^\Omega(x)} = \Omega(x)U_\mu(x)\Omega^\dagger(x+\hat\mu) \quad (13)$$

Because of the interplay between factors of $g$ in the expansions of $\Omega$ and $U_\mu$, the coefficient of $g^1$, which is the first non-trivial term, is the $\mathcal{O}(g^0)$ term in the gauge potential transformed to Laplacian gauge. Explicitly, it is

$$A_\mu^{\Omega(0)}(x) = \Xi^{(0)\dagger}\left[A_\mu(x) - il_1(x) + il_1(x+\hat\mu)\right]\Xi^{(0)} \quad (14)$$

As has been observed from the first discussions of the Laplacian gauge, the lattice divergence of this gauge potential vanishes

$$\Delta_\mu A_\mu^{\Omega(0)} = \Xi^{(0)\dagger}\left[\Delta_\mu A_\mu - i\Delta^{(0)}(-i\tilde\Delta^{(0)-1}\Delta_\mu A_\mu)\right]\Xi^{(0)} \quad (15)$$
$$= 0$$

The new result presented here, eliding much algebra, is that the next term in the expansion of the gauge-fixed gauge potential,

$$A_\mu^{\Omega(1)} = \frac{1}{2}\Xi^{(0)\dagger}\left\{[A_\mu, l_1 + l_1^{(+\hat\mu)}] - i[l_1, l_1^{(+\hat\mu)}]\right.$$
$$\left. -i(l_2 - l_2^\dagger) + i(l_2^{(+\hat\mu)} - l_2^{(+\hat\mu)\dagger})\right\}\Xi^{(0)} \quad (16)$$

where

$$l_i^{(+\hat\mu)}(x) \equiv l_i(x+\hat\mu) \quad (17)$$

also satisfies the Landau gauge condition

$$\Delta_\mu A_\mu^{\Omega(1)} = 0 \quad (18)$$

**4. NON-PERTURBATIVE FIXING**

The foregoing considerations can be verified in simulations. To do this, we applied the Laplacian gauge fixing procedure to ten ensembles of 20 well thermalized $6^4$ lattices generated with the simple Wilson action for the gauge group $SU(3)$ at values of $\beta = 6/g^2$ corresponding to $g = .1, .2, \ldots, .9, 1$. After fixing to Laplacian gauge, we evaluated the lattice divergence of the standard lattice gauge potential,

$$A_\mu(x) = \frac{[U_\mu(x) - U_\mu^\dagger(x)]_{traceless}}{2ig} \quad (19)$$

The average of $\Delta_\mu A_\mu$ over each ensemble is plotted vs the bare lattice coupling in Fig. 1.

Fig. 1 Deviation of Laplacian Gauge Fixed Lattices From Landau Gauge

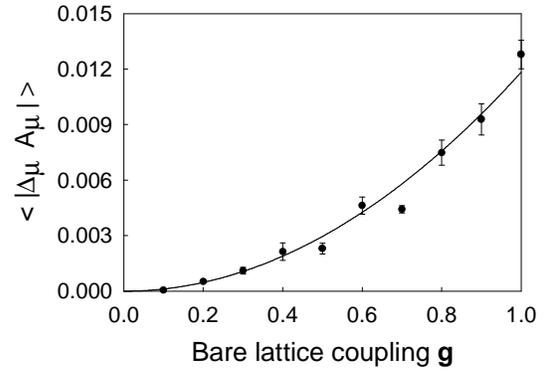

The curve is a quadratic fit to the square root of the ensemble average of $Tr(\Delta_\mu A_\mu)^2$ averaged over lattice sites.

The graph shows two relevant results. One is that the Landau gauge condition clearly fails to second order in perturbation theory. The other is that even at $g = 1$, which corresponds to $\beta = 6$, the magnitude of $\Delta_\mu A_\mu$ is only about 1% of its value on unfixed lattices.